\documentclass[twocolumn,aps,showpacs,prb,superscriptaddress]{revtex4-1}
\usepackage{natbib}
\usepackage{url}
\usepackage{bm}
\usepackage{textcase}
\usepackage{color}
\usepackage{graphicx}
\usepackage{amsmath}
\usepackage{ulem}
\usepackage[colorlinks=true,citecolor=blue]{hyperref} 
\hypersetup{pdftitle={Electronic structure and spin polarization in
    Fe(1-x)Mn(Co,Ni)(x)S(2) alloys}}

\hypersetup{pdfauthor={A. Houari and P.E. Bloechl}}
\hypersetup{pdfdisplaydoctitle}
\begin{document}
\title{Electronic structure and spin polarization of
  Fe$_{1-x}$Mn(Co,Ni)$_x$S$_2$ alloys from first principles}
\author{Abdesalem Houari}
\email[corresponding author:]{abdeslam.houari@univ-bejaia.dz}
\affiliation{Theoretical Physics Laboratory, 
             Department of Physics, 
             University of Bejaia, 
             Bejaia, Algeria}
\author{Peter E. Bl\"ochl}
\affiliation{Clausthal University of Technology, Institute for
Theoretical Physics, Leibnizstr.10, D-38678 Clausthal-Zellerfeld,
Germany}
\date{\today} 
\begin{abstract}
Alloying effects by T=Mn,Co,Ni-substitution on FeS$_2$ have been
investigated using density-functional calculations. The
alloys Fe$_{1-x}$T$_x$S$_2$ have been investigated for concentrations
$x=\frac{1}{4},\frac{1}{2},\frac{3}{4}$ together with the ground
states of the pure compounds. The electronic structure is discussed
with the goal to identify candidates for half metals, which are of
interest for spintronics applications. We find interesting candidates
at mean concentration of the Mn-doped FeS$_2$ and at low
concentrations for Ni-doped materials. For the Mn alloys we also note
the proximity to a low-spin to high spin transition. For Co-doped
materials we reproduce the well known finding of half metallicity over
the entire concentration range.
\end{abstract}
\pacs{71.10.-w, 71.15.Nc, 71.15.Mb}
%
\maketitle 
\section{Introduction}
\label {I}
The rapid development in the emerging field of {\it spintronics}, or
spin-electronics, has stimulated huge efforts to look for new
materials utilized for creating new electronic devices
\cite{bowen03_apl82_233,wolf01_science294_1488,kortright99_jmmm207_7,grynkewich04_mrs29_818,houari07_prb75_064420,houari08_cms43_392}.
One of the key issues in this area is to obtain materials with high
spin polarization at the Fermi level E$_F$, to be used as a source for
polarized spin injection. The best candidates for this purpose are the
so-called half-metallic ferromagnets (HMFs)
\cite{bowen03_apl82_233,fang02_jap91_8340,houari10_prb82_241201}. These
HMFs are metallic for one spin direction and insulating for the
opposite one, such offering a \textit{complete}, i.e. 100~\%, spin
polarization of the charge carriers.  However, it is first necessary
to have a good understanding of the underlying behavior in these
materials \cite{pask03_prb67_224420}.  Many investigations have
reported experimental evidence of either high or complete polarization
in some oxides and compounds (CrO$_2$
\cite{kamper87_prl59_2788,parker04_prb69_220413}, Fe$_3$O$_4$
\cite{parker04_prb69_220413,dedkov02_prb65_064417,morton02_sc513_L415}
and La$_{1-x}$Sr$_x$MnO$_3$
\cite{park98_prl81_1953,nadgorny01_prb63_184433}).

Pyrite-type disulfides have attracted considerable interest for
various reasons\cite{goodenough71_jssc3_26}. Semiconducting $ {\rm
  FeS_2} $ received widespread attention for its application in
photovoltaic energy conversion
\cite{eyert98_prb57_6350,ennaoui93_semsc29_289}. Substitution of Zn
for Fe in iron pyrite has thus been used to tune the optical band gap
in order to enhance the response to the solar spectrum
\cite{buker99_jes1_261}.

This class of materials has a variety of interesting
properties\cite{temmerman93_pcm20_249}.  While $ {\rm FeS_2} $ is a
van Vleck paramagnet, metallic $ {\rm CoS_2} $ displays long-range
ferromagnetic order\cite{hobbs99_jpc11_8197}. 

In contrast, ${\rm MnS_2}$ is found to be an antiferromagnetic insulator 
in the ground state, with a transition temperature of T$_N$=48.2 K.
\cite{hastings76_prb14_1995,chattopadhyay84_ssc50_865,kahn83_jpc16_4011}
The insulating behavior is also found in $ {\rm  NiS_2} $, but with a
disagreement about the ground state magnetic order. While some reports 
describe $ {\rm NiS_2} $ as an antiferromagnet\cite{krill76_jpc9_761,matsuura96_prb53_7584,yao96_prb54_17469},
some others found it as a paramagnet \cite{kautz72_prb6_2078,kikuchi78_jpsj44_410,li74_prl3_470}.
The insulating behavior of ${\rm MnS_2}$ and ${\rm NiS_2}$ has been attributed to the presence of
strong electronic correlations\cite{bither68_ic7_2208,
adachi72_jpsj32_573, ogawa79_jap50_2308, hastings70_ibmjrd14_227,
miyadai77_pbc86_901, chattopadhyay91_prb44_7394,
bocquet96_jpcm8_2389, hiraka98_jmmm177_1349}.
Some resonant photoemission studies\cite{fujimori96_prb54_16329}, however, suggested
that $ {\rm NiS_2} $ is a {\it p}-to-{\it d} charge-transfer insulator
rather than a {\it d-d} Mott-Hubbard insulator, with a gap between the
occupied S {\it 3p}-states and the empty Ni {\it 3d}-states.

Among the pyrite-type disulfides, $ {\rm Fe_{1-x}Co_xS_2} $ has drawn
much attention for spintronic applications. It is considered as a
model for band gap tuning and for studying half-metallicity
\cite{wang04_prb69_094412}. It has been shown that the spin
polarization in this system can be controlled with respect to alloying
and should not be sensitive to crystallographic disorder
\cite{mazin00_apl77_3000}.  Many experimental investigations and
transport measurements have been carried
out\cite{ramesha04_prb70_214409,
  wang05_prl94_056602,wang06_apl88_232509, cheng03_jap93_6847}.  Wang
{\it et al} \cite{wang05_prl94_056602} have recently combined indirect
transport probes with a direct measurement by PCAR (point contact
Andr\'eev reflection), to prove that the spin polarization can be
continuously tuned over a wide range.

Moreover, several {\it ab-initio} calculations have been reported on
the electronic structure of the $ {\rm Fe_{1-x}Co_xS_2} $ alloys
\cite{zhao93_prb48_15781, mazin00_apl77_3000, umemoto06_pssb243_2117,
  leighton07_jpcm19_315219}.  Zhao {\it et al}
\cite{zhao93_prb48_15781} predicted already in the early nineties on
the basis of the Linear Combination of Atomic Orbital (LCAO) method
that half-metallicity might be obtained when alloying $ {\rm FeS_2} $
with $ {\rm CoS_2} $. A very important contribution has been the work
by Mazin {\it et al} \cite{mazin00_apl77_3000, umemoto06_pssb243_2117}
who performed detailed calculations using the Linear Muffin Tin
Orbital (LMTO) method as well as the full-potential Linear Augmented
Plane Wave (FP-LAPW) method. This work showed clearly that
half-metallic ferromagnetism is present in the region $0.25 \leq x
\leq 0.85$ , with magnetic moments in a very good agreement with
experiment.

Besides $ {\rm Fe_{1-x}Co_xS_2} $, some
attempts\cite{sun11_prb84_245211} have been made to improve the band
gap of $ {\rm FeS_2} $. Different types of substitution have been
tested ($ {\rm Fe_{1-x}T_xS_2} $, where $ {\rm T = Be, Mg, Ca, Os, Ru}
$ ...etc) in order to increase the gap for technical
applications.

However, little attention has been
paid to the $ {\rm Fe_{1-x}(Mn,Ni)_xS_2} $ systems, even-though the
pure compounds (Mn and Ni disulfides) share the same pyrite crystal
structure.  In early experiments, $ {\rm Fe_{1-x}Ni_xS_2}$ solid
solution has been synthesized
\cite{bither68_ic7_2208,bouchard68_mrb3_563,bither70_jssc1_526}.

Dilute substitution (less than 2\%) of Fe in $ {\rm MnS_2} $
has been reported in experimental studies, using M\"ossbauer technique
\cite {bargeron71_ic10_1338,kahn83_jpc16_4011}. Recently, 
Persson {\it et al} \cite {persson06_prb73_115201}
have investigated theoretically in the framework of GGA+U the spin transitions in 
$ {\rm Fe_xMn_{1-x}S_2} $. It has been found that the spin transition
pressure decreases with growing Fe substitution in $ {\rm MnS_2} $.

In the present work, we investigate the $ {\rm Fe_{1-x}(Mn,Ni)_xS_2} $
systems, in order to get a detailed picture of the ferromagnetic
trends in the whole series, and to provide further insight into the
mechanism of half-metallicity and spin polarization in these systems.
The paper is organized as follows: In Sec.~\ref{paw}, we describe the
theoretical method and the computational details.  A brief summary of
the crystal structure is given in Sec~\ref{str}. In Sec.~\ref{results}
we present our results on the three compounds $ {\rm
  Fe_{1-x}(Mn,Co,Ni)_xS_2} $.  We summarize in Sec.~\ref{conclusion}
     
\section{Computational Method}
\label{paw}
In the present study, we performed first-principles calculations in
the framework of density-functional theory (DFT)
\cite{hohenberg64_pr136_B864, kohn65_pr140_1133}.  The
exchange-correlation effects were taken into account within the
generalized gradient approximation (GGA) of Perdew-Burke-Ernzerhof
\cite{perdew96_prl77_3865}. We also performed some calculations using
the PBE0r functional, which replaces a fraction of the exchange term
of the PBE functional with the exact
Fock-term\cite{becke93_jcp98_1372,perdew96_jcp105_9982}. Unlike the
PBE0\cite{adamo99_jcp110_6158} hybrid functional, we follow the idea
of range separated hybrid functionals\cite{heyd03_jcp118_8207} and
restrict this correction in the PBE0r functional to the onsite
interactions in a local orbital basis.

The calculations have been performed with the Projector Augmented Wave
(PAW) method\cite{bloechl94_prb50_17953} as implemented in the CP-PAW
code.  The PAW method is an {\it all-electron} electronic structure
method.  The CP-PAW code is the original implementation of the PAW
method and employs the framework of {\it ab-initio} molecular dynamics
(AIMD)\cite{car85_prl55_2471} for the optimization of wave functions
and atomic structure.  For the augmentation, we used a $s^1p^1d^1$ set
of projector functions for all atoms, where the superscripts denotes
the number of projector functions angular momentum channel.  A plane
wave cutoff of 30~Ry has been chosen for the wave functions and 60~Ry
for the charge density.

The Brillouin-zone integration has been performed with the linear
tetrahedron method\cite{jepsen71_ssc9_1763, lehmann72_pssb54_469} and
the so-called Bl\"ochl corrections\cite{bloechl94_prb49_16223}. A
$5\times5\times5$ {\it k}-point mesh has been used for the
simple-cubic unit cell containing four transition-metal ions.

All structural parameters including the lattice parameters have been
optimized.

\section{Crystal Structures}
\label{str}
\begin{figure}[htbp]
\begin{center}
\includegraphics[width=0.48\linewidth]{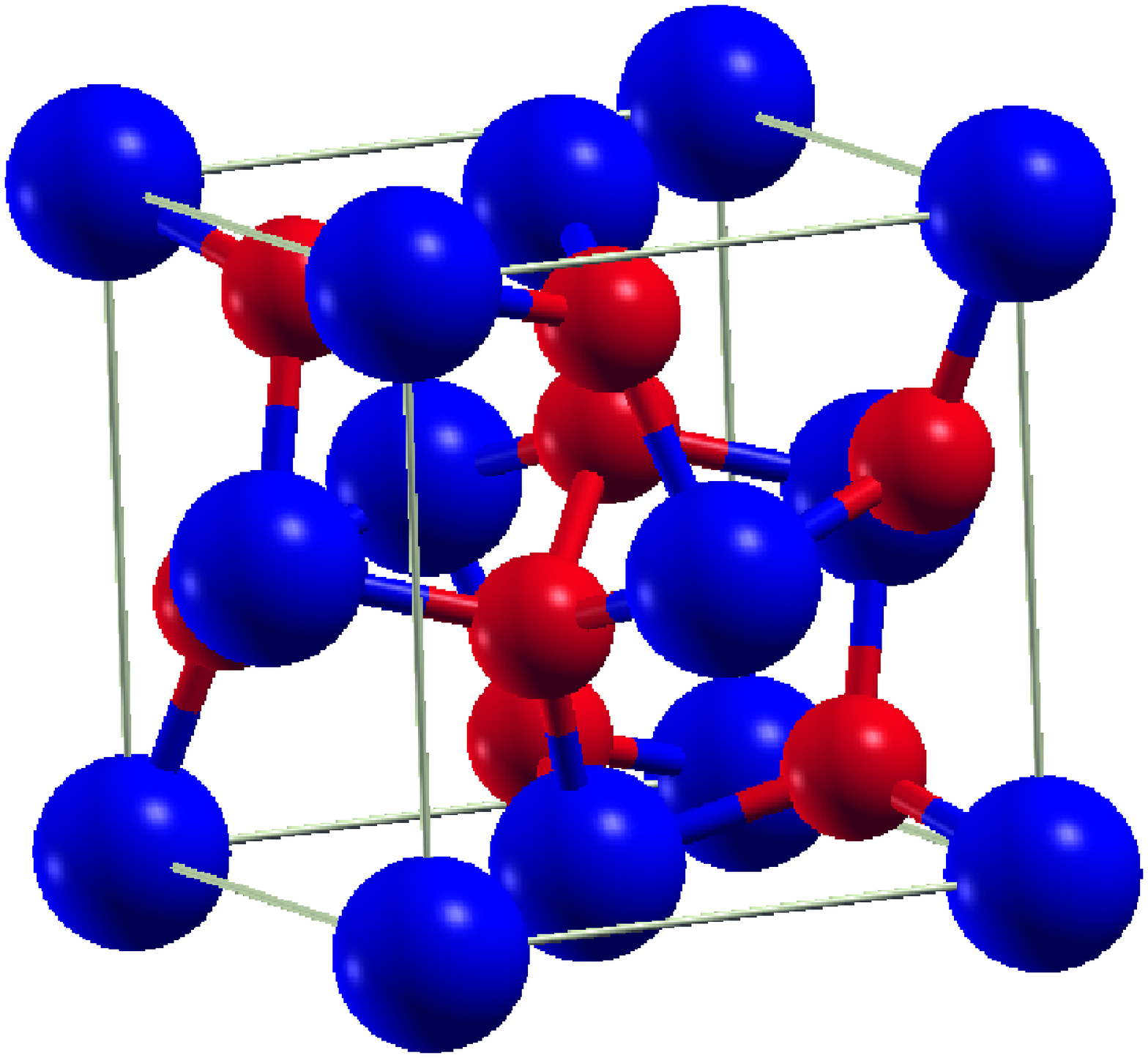} 
\includegraphics[width=0.48\linewidth]{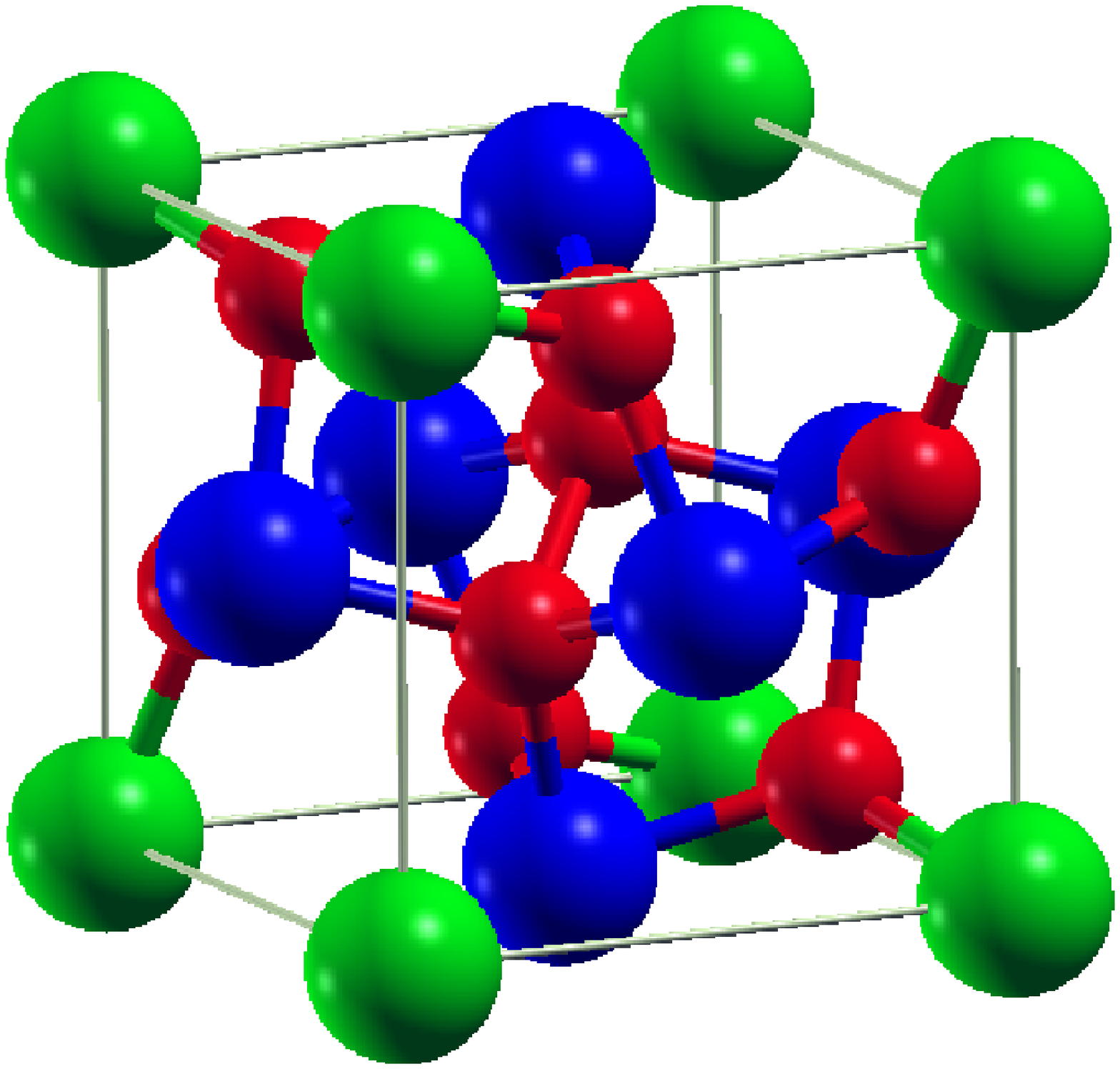}
\end{center}
\caption{\label{fig:pyrite} (Color online) Crystal structures of the
  pure iron pyrite $ {\rm FeS_2} $ (left) and the substituted compound
  $ {\rm Fe_{\frac{3}{4}}Mn_{\frac{1}{4}}S_2} $ (right). Fe, Mn and S
  atoms are printed in blue, green and red, respectively.
}
\end{figure}
In this section, we describe the crystal structures used in our
calculations. Many transition-metal disulfides $ {\rm TS_2}$ (with T =
Mn, Fe, Co, Ni, Cu, Ru and Os ... etc), crystallize in the cubic
pyrite structure. The pyrite structure is based on a simple cubic
({\it sc}) lattice with space group Pa$\Bar{3}(T_h^6$)
\cite{finklea76_acsa32_529,stevens80_ic19_813,eyert98_prb57_6350}. The
transition-metal atoms occupy the Wyckoff positions (4a) and sulfur
atoms occupy the positions (8c). The $ {\rm FeS_2}$ pyrite crystal
structure is displayed in Fig.\ref{fig:pyrite}, as well as an
illustration of one alloyed compound $ {\rm
  Fe_{\frac{3}{4}}Mn_{\frac{1}{4}}S_2} $.

The pyrite crystal structure can be rationalized in terms of the {\it
  fcc}-NaCl structure with the sublattices occupied by
transition-metal ions atoms and sulfur dimers, respectively. The
sulfur dumbbells are oriented along the $\langle 111\rangle$ axes.
Because the distinct dumbbells are oriented in all four distinct
$\langle111\rangle$ directions, the simple-cubic unit cell with four
formula units of FeS$_2$ is employed.

Whereas the sulfur atoms are tetrahedrally coordinated by one sulfur
and three iron atoms, the six nearest-neighbor sulfur atoms at each
metal site form slightly distorted octahedra.

\begin{table*}[t]
\caption{\label{tab:table1} Calculated data lattice constants {\it a}
  (in \AA), internal parameter {\it u} and averaged bond lengths of
  metal-sulfur and sulfur-sulfur bonds. Experimental
  data\cite{fujimori96_prb54_16329} are shown in parentheses. }
\begin{ruledtabular}
\begin{tabular}{llllll}
$ {\rm {Fe_{1-x}Mn_xS_2}} $ & $ a$[\AA] &  $ u$  & ${\rm {d_{Fe-S}}}$[\AA] 
&${\rm {d_{Mn-S}}}$[\AA] & ${\rm {d_{S-S}}}$[\AA] \\
\hline
$ {\rm x = 0.00} $                    & 5.405 (5.416) &0.387 (0.385)&2.263          &        & 2.160  \\
$ {\rm x = 0.25} $                    & 5.432         &             &2.257          &2.272   & 2.274  \\
$ {\rm x = 0.50} $                    & 5.648         &             &2.270          & 2.293  & 2.209  \\
$ {\rm x = 0.75} $                    & 5.832         &             &2.411          & 2.432 & 2.411  \\
$ {\rm x = 1.00} $                    & 6.062(6.091)  &0.402 (0.409)&               & 2.527 & 2.620  \\
\hline
$ {\rm {Fe_{1-x}Co_xS_2}} $ &         &    &          &${\rm {d_{Co-S}}}$ &                      \\
\hline
$ {\rm x = 0.00} $                    &5.405(5.416)   & 0.385 (0.385)&2.263          &       & 2.160 \\
$ {\rm x = 0.25} $                    &5.420          &              &2.249  & 2.285 & 2.181   \\
$ {\rm x = 0.50} $                    &5.440          &              & 2.260 & 2.292 &2.145   \\
$ {\rm x = 0.75} $                    &5.471          &              &2.264  &2.296  & 2.125      \\
$ {\rm x = 1.00} $                    &5.489 (5.539)  &0.388 (0.390) &2.310  &       &2.101\\
\hline
$ {\rm {Fe_{1-x}Ni_xS_2}} $ &         &    &           &${\rm {d_{Ni-S}}}$ &  \\
\hline
$ {\rm x = 0.00} $                    & 5.405 (5.416) &0.385 (0.385)&2.263         &          &   2.160   \\
$ {\rm x = 0.25} $                    & 5.445         &             &2.247 & 2.340 &2.146\\
$ {\rm x = 0.50} $                    & 5.494         &             &2.259 &2.331 & 2.090  \\
$ {\rm x = 0.75} $                    & 5.542         &             &2.249 &2.325 & 2.042 \\
$ {\rm x = 1.00} $                    & 5.572 (5.620) &0.393 (0.394)& 2.359&      &2.064  \\ 
\end{tabular}
\end{ruledtabular}
\end{table*}

The pyrite structure is determined by the lattice constant and an
internal parameter $u$.  The internal parameter $u$ and the lattice
constant determine the bond length of the sulfur dumbbell as
$d_{S-S}=\sqrt{3}(1-2u)a_{lat}$ and the Fe-S distance is
$d_{Fe-S}=\sqrt{2(1-u)^2+u^2}a_{lat}$.\cite{birkholz08_pssb245_1858}

The crystal structure of the pure disulfides is well established
experimentally as well as theoretically (experimental data are included in 
table \ref{tab:table1}). No structural information has been reported, 
however, for the alloyed compounds.

Calculated structural parameters and the lattice constants for the
complete series are given in Table \ref{tab:table1}.
The bond lengths are averaged. Individual
bond lengths differ by 0.2~\%. from these averages.
For the pure pure disulfides of Fe, Mn, Co, and Ni, the calculated bond lengths
have been compared to experimental data\cite{fujimori96_prb54_16329}
and found to deviate less than 0.5~\%.

We considered those ordered alloy structures, that can be described
 within the simple cubic unit cell of pyrite. For each of the concentrations 
$x=\frac{1}{4}$, $x=\frac{1}{2}$, and $x=\frac{3}{4}$ only one symmetry 
inequivalent arrangement of the transition-metal sites exists. As seen in 
Fig.1 the minority alloying element for the occupations $x=\frac{1}{4}$ and 
$x=\frac{3}{4}$ occupies the cube corners, while the face centers are occupied 
with the majority element. For $x=\frac{1}{2}$, the alloying elements occupy 
alternating $\langle001\rangle$ planes.

\section{Results and Discussions}
\label{results}
The main goal of the present paper is the understanding of the effect
of alloying $ {\rm FeS_2}$ with Mn, Co and Ni on the electronic and
magnetic properties.  In the following, we investigate the electronic
properties of the three different alloys $ {\rm
  Fe_{1-x}Mn(Co,Ni)_xS_2}$, and discuss the findings into the context
of the pure compounds.
    
\subsection{$ {\rm \bf Fe_{1-x}Mn_xS_2} $}
\label{mnfes2}
\begin{figure}[htbp]
\includegraphics[width=\linewidth]{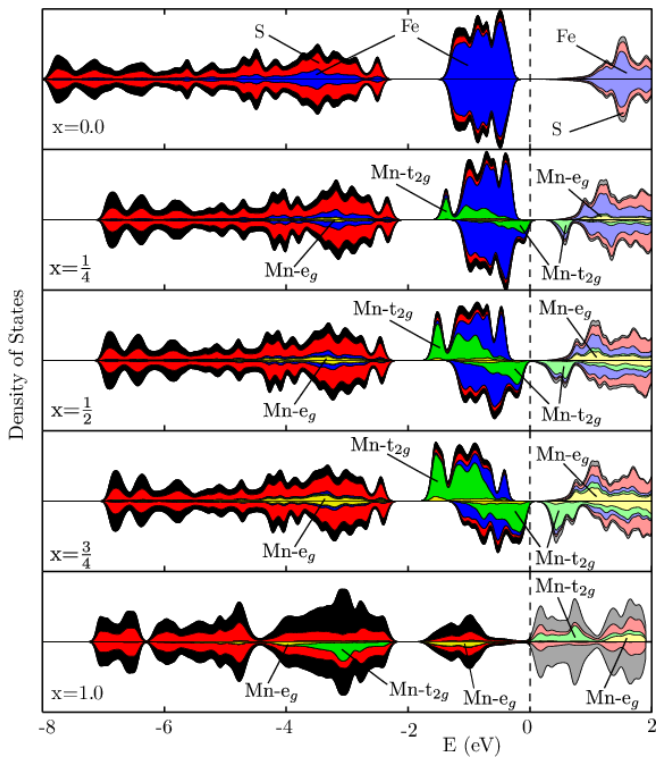}
\caption{(Color online) Density of states (DoS) of
  Fe$_{1-x}$Mn$_x$S$_2$ alloys for $x=0, \frac{1}{4}, \frac{1}{2},
  \frac{3}{4}$ and for $x=1$. The origin of the energy axis is set to
  the Fermi level $E_F$, which is indicated by the dashed line. The
  DoS of spin-down electrons is drawn with the axis pointing
  downward. The total density of states is given by the black
  envelope. The projected DoS are drawn as distinct colored areas
  on top of the total DoS.  The DoS of the Fe-d states are represented
  by the blue area and the DoS of the sulfur states by the red area.
  The dopant (Mn) {\it 3d} states are divided into t$_{2g}$ states
  shown in green and e$_g$ states shown in yellow.  For the
  antiferromagnetic MnS$_2$ the projected DoS of the d-states is shown
  only for those Mn with the majority-spin direction in the down
  channel.  }
\label{fig:femn-dos}
\end{figure}

The density of states (DOS) of $ {\rm Fe_{1-x}Mn_xS_2} $ at $\rm x =
0, \frac{1}{4}, \frac{1}{2} , \frac{3}{4}$ and 1 are shown in
Fig.\ref{fig:femn-dos} .

The electronic structure of pure FeS$_2$ can be rationalized as
follows: The sulfur dumbbells form negatively charged $S_2^{2-}$ ions
with all valence states filled except for the antibonding $\sigma^*$
orbital between the sulfur ions. Consequently, each iron atom has an
oxidation state $2+$. The iron atom exists in a low-spin occupation
with filled t$_{2g}$ orbitals and empty e$_g$ orbitals. Between the
t$_{2g}$ and the $e_g$ states exists a band gap of $0.69$~eV in our
calculations and $\sim 0.9$~eV in photoconductivity
measurements\cite{ennaoui93_semsc29_289,schlegel76_jpc9_3363,li74_prl3_470}. The
underestimation is common for GGA calculations. It should be noted
that the band gap in our calculations is determined by a conduction
band state with sulfur character that extends with a very small
density of states below the d-states.

The density of states of FeS$_2$ seen in the top graph of
Fig.~\ref{fig:femn-dos} is dominated by a the sulfur $p$ states
forming a lower valence band extending from -8~eV to -2~eV.  The
sulfur $s$-states lie below the $p$-states and are outside the energy
region shown. Above the sulfur states the Fe-t$_{2g}$ states form an
isolated narrow band. Thus the valence-band edge is of Fe-t$_{2g}$
character. The conduction band is actually a tail of
predominantly sulfur character, which extends below the prominent
Fe-e$_g$ states, which rise up at about +1~eV. Due to the small
density of states the tail of the sulfur states is easily overlooked
and it is hardly visible in the figure.

At first, we investigated the concentration dependence on the 
basis of ferromagnetically aligned Mn-ions. This is the case 
of technological interest, because the spins may be aligned by 
an external magnetic field. The corresponding results are shown 
in Fig.~\ref{fig:femn-dos}. The relative stability of the 
ferromagnetic order relative to antiferromagnetic states will 
be addressed later in this section.

The Mn-t$_{2g}$ states are located in the same energy region as the
Fe-t$_{2g}$ states. Because the Mn ions have one electron less than
the Fe ions, the Mn-t$_{2g}$ states have a hole in the t$_{2g}$
shell. This spin of this hole results in a magnetic moment, so that
the majority-spin and minority-spin states of Mn-t$_{2g}$ character
are shifted against each other. The top of the valence band in the
majority-spin direction is of Fe character while that of the
minority-spin direction is of Mn character.

Substitutional alloying Mn for Fe introduces empty states into the
gap between t$_{2g}$ and e$_g$ states of FeS$_2$. These states derive
from a Mn-t$_{2g}$ orbitals. The Mn ions exist in a low-spin $2+$
oxidation state, for which the e$_g$ orbitals are empty and the
t$_{2g}$ are filled except for one orbital, which appears in the band
gap. 

The majority-spin direction has a band gap with only a weak
concentration dependence. This gap decreases abruptly from 0.69~eV
in FeS$_2$ to 0.37-0.41~eV in the alloys.  In contrast,
the band gap in the minority-spin direction shrinks with increasing
concentration as the empty t$_{2g}$ states hybridize more effectively.

For low concentration, i.e. for $x=0.25$, a semiconductor is
obtained with a band gap of 0.35~eV.  Valence and conduction band are
both in the minority-spin direction.  The electron transport
properties are thus dominated by a single spin direction.

At larger Mn concentration, that is for $x=\frac{1}{2}$ and
$x=\frac{3}{4}$, the minority-spin band gap almost collapses with
E$_g$ below 0.01~eV. Because this band gap is in the minority-spin
direction, this material can be considered a half metal for practical
purposes. We expect this material to be of interest for spintronics
applications.

The Fermi-level moves, however, at higher concentration close to the
conduction-band bottom of the majority-spin direction.  This is not
immediately evident from Fig.~\ref{fig:femn-dos} because of the
low density of states in the conduction band of the majority-spin
direction.

After having discussed the electronic structure of the alloys in
  the ferromagnetic state, we investigated their stability relative to a
  competing antiferromagnetic phase. We investigated the type-III
  single-k antiferromagnetic ordering, which has been established
  experimentally\cite{chattopadhyay91_prb44_7394} for MnS$_2$, as well
  as the order with antiferromagneticly coupled (111) planes. The former
  requires to double the unit cell in $\langle100\rangle$ direction.

Both antiferromagnetic arrangements produce analogous results:
  The ferromagnetic order is stable for low Mn concentration,
  $x=\frac{1}{4}$ and $x=\frac{1}{2}$, while at higher Mn
  concentration $x=\frac{3}{4}$ and $x=1$ the antiferromagnetic order
  is favorable.

The antiferromagnetic materials are described as
  metals.  The metallic behavior of the antiferromagnets is attributed
  to a deficiency of the density functionals used.  Calculations with
  hybrid functionals produce an insulating ground state.

Thus, our calculations predict half-metallic behavior in the
  half-doped regime. At lower concentration the material becomes more
  insulating, while at higher doping, that is between $x=\frac{1}{2}$
  and $x=\frac{3}{4}$, the material turns antiferromagnetic. It may be
  possible to stabilize the ferromagnetic order also for higher
  concentrations through an external magnetic field.

\begin{figure}[htbp]
\includegraphics[width=\linewidth]{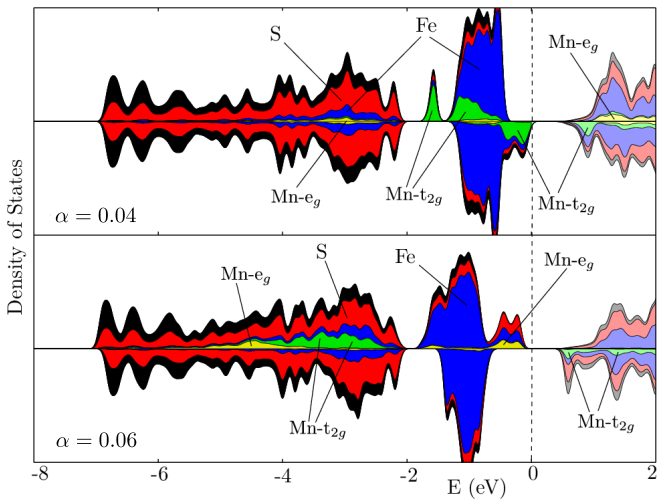}
\caption{(Color online) Density of states (DoS) of
  Fe$_{\frac{3}{4}}$Mn$_{\frac{1}{4}}$S$_2$ alloy, in the low spin
  (upper panel) and high spin (lower panel) states of Mn, using hybrid
  functionals. Two different weights of the Fock term,
  ($\alpha$=0.04 and $\alpha$=0.06), have been used to direct the system
  into the respective configuration.  Further description as in
  Fig.~\ref{fig:femn-dos}.}
\label{fig:femn-hyb-dos}
\end{figure}

The low-spin character of the Mn-ions in the alloys differs from the
high-spin character of the pure MnS$_2$. Even in the GGA calculations,
that is without the Fock term, we find the pure compound to be a
high-spin antiferromagnet, in agreement with recent GGA
calculations\cite{persson06_prb73_115201}.  Experimentally, MnS$_2$ is
found to be an antiferromagnetic
insulator\cite{chattopadhyay91_prb44_7394}.  The calculated band gap
within GGA vanishes, while the local hybrid functionals with an
admixture of 10~\% of Fock exchange provides a band gap of 0.91~eV in
agreement with experiment of 1~eV\cite{brostigen70_actachem24_2993}.

Performing calculations with a local hybrid functional for the alloys,
we noticed that the the system is close to a transition between low-spin
and to high-spin Mn states. Replacing more than 5\% of the PBE
exchange by the local Fock term drives the Mn ions into a high-spin
state with a filled d-shell in the majority-spin direction and an
empty d-shell in the minority-spin direction. In this configuration that
compound is an insulator. Using hybrid calculations with a
Fock-exchange admixture of 10\%, we obtain the high-spin configuration
of Mn ions for all concentrations considered.

Similar results have been obtained by Persson {\it et al} \cite
{persson06_prb73_115201} using the GGA+U method within the
implementation of Liechtenstein {\it et al}
\cite{liechtenstein95_prb52_R5467}.  They have shown the existence of
high-spin to low-spin transitions for $x=\frac{1}{4}$, $x=\frac{1}{2}$
and $x=\frac{3}{4}$, and they calculated the corresponding pressure of
transition.

To illustrate this transition between low-spin and to high-spin, we
show in Fig.~\ref{fig:femn-hyb-dos} the density of states for low-spin
and high-spin for $x=\frac{1}{4}$. We used hybrid functionals with two
Fock term weights, namely $\alpha=0.04$ and $\alpha=0.06$, to show 
the main changes due to the
transition. The different values have been chosen to guide the system
into the chosen configuration.

In the low-spin configurations that Mn-t$_{2g}$ states are filled
except for one hole in the minority-spin direction. The e$_g$ states
on the other hand are empty. In the high-spin configuration, all
majority-spin d-states on Mn are occupied and all minority spin
d-states are empty.

\subsection{$ {\rm \bf Fe_{1-x}Co_xS_2} $}
\label{cofes2}
Let us now turn to the alloy $ {\rm Fe_{1-x}Co_xS_2} $.  It has been
studied in detail, both experimentally\cite{leighton07_jpcm19_315219,
  wang05_prl94_056602} and theoretically\cite{mazin00_apl77_3000,
  umemoto06_pssb243_2117}, and our results are consistent with
previous ones.  They are added here for the sake of consistency and
completeness.
\begin{figure}[htbp]
\includegraphics[width=\linewidth]{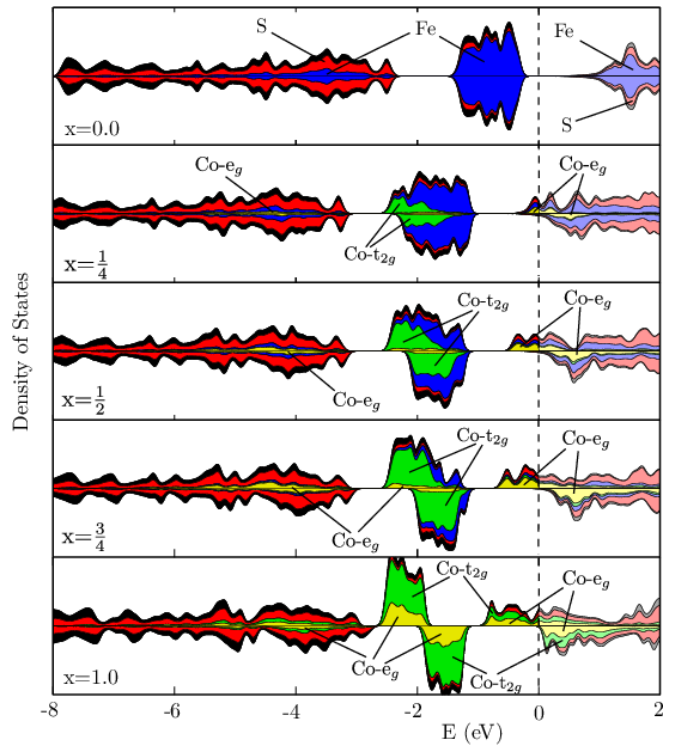}
\caption{(Color online) Density of states (DoS) of
  Fe$_{1-x}$Co$_x$S$_2$ alloys for $x=0, \frac{1}{4}, \frac{1}{2},
  \frac{3}{4}$ and for $x=1$.  Further description as in
  Fig.~\ref{fig:femn-dos}.}
\label{fig:feco-dos}
\end{figure}

The density of states of the series with x = 0, $\frac{1}{4}$,
$\frac{1}{2}$, $\frac{3}{4}$ and 1 is given in Fig.\ref{fig:feco-dos}.
It agrees with DFT calculations by Umemoto {\it et al}
\cite{umemoto06_pssb243_2117}. The calculated electronic structure
agrees also well with other theoretical and experimental data found in
literature \cite{leighton07_jpcm19_315219, wang05_prl94_056602}.

The Co$^{2+}$ ions have a filled t$_{2g}$ shell and one electron in
the e$_g$ shell. The Fermi level is pinned in the
e$_g$ states of Co of the majority-spin direction. As a consequence
the alloys with Co are half metals.

The Co-$t_{2g}$ states are located in the same energy
region as the Fe-$t_{2g}$ states, i.e. in the upper part of the
valence band between -3 eV and -1 eV, for all the alloys.  The
Co-$e_g$ bands form the lower part of the conduction band.  However,
some weight is also present in the top region of the occupied sulfur
$p$-band between -5.5 eV and -3.5 eV.  The spin splitting of the
Co-$t_{2g}$ states is smaller than that of the Mn-ions in the
Fe$_{1-x}$Mn$_x$S$_2$ alloys. 

\subsection{$ {\rm \bf Fe_{1-x}Ni_xS_2} $ }
\label{nifes2}
Early experimental studies\cite{bither70_jssc1_526} investigated the $
{\rm Fe_{1-x}Ni_xS_2} $ alloys for concentrations $0.4 \leq$ x $\leq
0.6$. In this range of substitution, their magnetic measurements have
shown that $ {\rm Fe_{1-x}Ni_xS_2} $ are paramagnetic metals.

The alloying effect of FeS$_2$ with Ni can be understood again from
the the point of view of Ni$^{2+}$ ions. Ni$^{2+}$ has a half-filled
e$_g$ shell being occupied with two electrons. The near degeneracy of
e$_g$ states is lifted by Hund's rule splitting, which concentrates
both e$_g$ electrons in one spin direction which results in magnetic
ions.

\begin{figure}[htbp]
\includegraphics[width=\linewidth]{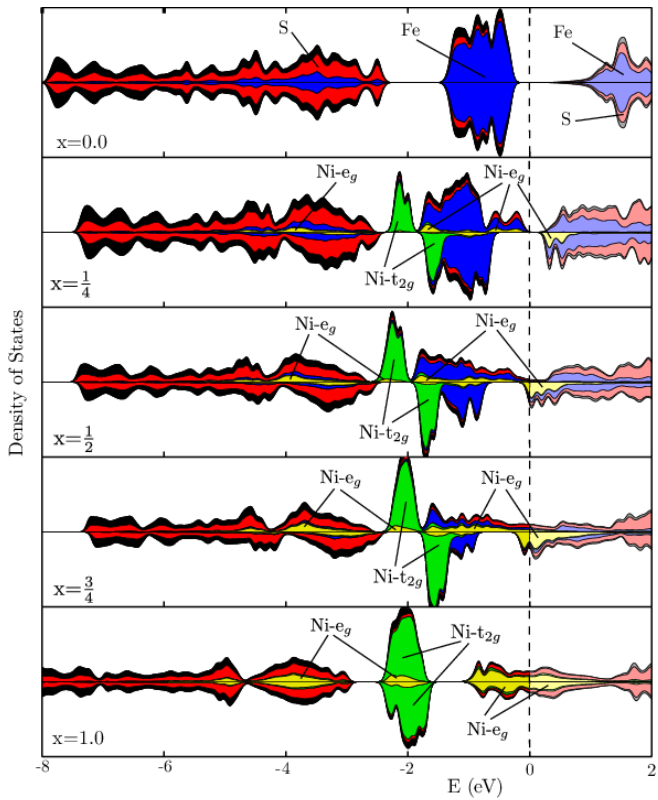}
\caption{(Color online) Density of states (DoS) of
  Fe$_{1-x}$Ni$_x$S$_2$ alloys for $x=0, \frac{1}{4}, \frac{1}{2},
  \frac{3}{4}$ and for $x=1$. The origin of the energy axis is set to
  E$_F$, which is indicated by the dashed line. The total density of
  states is given by the black envelope. The projected density of
  states are drawn as distinct colored areas on top of the total
  density of states.  The Fe DoS are represented by the blue area and
  the sulfur DoS by the red area.  The Ni-{\it 3d} states are divided
  into t$_{2g}$ states shown in green and e$_g$ states shown in
  yellow}
\label{fig:feni-dos}
\end{figure}

For low concentration, i.e. for $x=\frac{1}{4}$, we find a conduction
band in the minority-spin direction of predominantly Ni-e$_g$
character. It is separated by a band gap of 0.77~eV from the
Fe-t$_{2g}$ states. In the majority-spin direction, we find filled Ni
e$_{g}$ states below the Fermi level that are strongly hybridized with
sulfur and Fe-t$_2g$ states. In the majority-spin direction valence
and conduction bands overlap, which turns this material into a
half-metal.

At higher concentration, i.e. for $x=\frac{1}{2},\frac{3}{4}$, the
Fermi level shifts into the formerly empty Ni-e$_g$ band of the
minority spin direction. These electrons are removed from the states
that derive from the majority-spin Ni-e$_g$ states.

 As a consequence, the magnetic moment decreases with increasing
 concentration from 2~$\mu_B$ per Ni atom for $x=0$ to $1.7~\mu_B$ per
 Ni atom for $x=\frac{1}{2}$ and to $1.3~\mu_B$ per Ni atom for
 $x=\frac{3}{4}$. 

In agreement with the experimental magnetic ordering
\cite{yao96_prb54_17469}, pure NiS$_2$ is found to be an
antiferromagnet.  However, the obtained metallic character disagrees
with experiments\cite{kautz72_prb6_2078}, which find a semiconducting
ground state with an indirect band gap of 0.27~eV. This metallic
behavior is attributed to the well-known band gap problem of DFT To
resolve this issue, we carried out local hybrid functional
calculations and found an NiS$_2$ semiconductor with a gap of $\sim$
0.4 eV consistent with experiment.

In view of the magnetic results of NiS$_2$ , we explored
  the stability of the Fe$_{1-x}$Ni$_x$S$_2$ alloys against a
  transition to an antiferromagentic order. We explored the stability
  of the same two antiferrmagnetic orders used before for the Mn-Fe
  alloys. In contrast to the Mn-Fe system, however, the Ni-Fe alloys remain
  ferromagnetic for all alloys studied $x=\frac{1}{4},\frac{1}{2}$ and
  $\frac{3}{2}$.

 Comparing the alloys with Mn, Co, and Ni, we observe the trend that
 the d-bands shift down in energy with increasing atomic number due to
 the Coulomb attraction to the increased nuclear charge.

\section{Summary and Conclusion}
\label{conclusion}
Using first-principles calculations, we explored the possibility of
obtaining half metals by alloying the transition metals Mn,Co, and Ni
to FeS$_2$.  Alloys with dopant concentration
$x=\frac{1}{4},\frac{1}{2}$ and $x=\frac{3}{4}$ have been investigated
together with the pure compounds. Having the application on
spintronics in mind, this work focusses on the ferromagnetic states.
For the $ {\rm Fe_{1-x}Co_xS_2} $ alloys, we confirm
previous experimental and theoretical results. We provide indications
which alloys are promising candidates for half metals.

For the Mn doped alloys, we obtain a semiconductor at low
concentration. At higher concentrations this gap closes in the
minority-spin direction for all practical purposes. Thus we expect
this material to be suitable as a half-metal. It should be noted,
however, that the material is close to a low-spin to high-spin
transition, which can be triggered by changing the density functional
to hybrid functionals. In the high-spin configuration we obtain the
material to be insulating.

For the $ {\rm Fe_{1-x}Co_xS_2} $ alloy, half-metallic ferromagnets 
are obtained for the full range of concentrations investigated. The system

For the Ni-doped alloys we obtain a half metal for low doping,
i.e. for $x=\frac{1}{4}$. At higher doping the system turns into a
full metal.

\begin{acknowledgements}
A.H. gratefully acknowledges financial support by Bejaia university.
Financial support by the Deutsche Forschungsgemeinschaft through FOR
1346 is gratefully acknowledged.
\end{acknowledgements}

\end{document}